A microwave interferometer of the Michelson-type to improve the dynamic range of broadband ferromagnetic resonance measurements

E. R. J. Edwards[1*], A. B. Kos[1], M. Weiler[2,3], T. J. Silva[1]

1- Quantum Electromagnetics Division, National Institute of Standards and Technology, Boulder, CO 80305
2- Walther-Meißner-Institut, Bayerische Akademie der Wissenschaften, 85748 Garching, Germany
3- Physik-Department, Technische Universität München, 85748 Garching, Germany

Abstract

We present a Michelson-type microwave interferometer for use in ferromagnetic resonance experiments. The interferometer is capable of broadband operation without manual adjustment of phase delay or amplitude attenuation. A prototype of the design shows significant improvement of the signal-to-noise ratio when compared to non-interferometric ferromagnetic resonance experiments. We demonstrate that this increase in sensitivity can lead to a drastic increase in the data acquisition rate for hard-to-measure thin films that otherwise would require long integration times.

Introduction

Vector network analyzer-based ferromagnetic resonance (VNA-FMR) is a well-established technique for the characterization of magnetic thin films [Kalarickal 2006, Maksymov 2014, Silva 2016]. In VNA-FMR, a magnetic thin film is loaded onto a planar transmission line that is then placed in an external magnetic field. A VNA is used to measure the S-parameters of the loaded transmission line as a function of external field and frequency. The high sensitivity of VNA-FMR [Neudecker 2006] derives from the large dynamic range of commercially available VNAs, usually at least 120 dB in a 10 Hz measurement bandwidth at microwave frequencies. At the same time, such a large dynamic range is required due to the small filling factors [Pozar 2011] for this measurement geometry; most of the energy of the driving microwave magnetic field does not interact with the magnetic thin film.

While it is possible to measure and extract important material information from a wide variety of samples despite the small filling factor [Maksymov 2014, Shaw 2011, Shaw 2013, Boone 2013] the presence of a large background signal at the VNA receiver introduces a large offset as well as background noise. The offset decreases the dynamic range available for signal detection, while the background noise slows down the measurement. Eliminating this background signal during measurement would both increase the available dynamic range of the setup and decrease acquisition times.

---

In fact, it has long been known that interferometric techniques can be used to realize a strong suppression of the background signal in magnetic resonance experiments [Poole 1983]. Recently, several works have reported large increases in the sensitivity of VNA-FMR measurements using these techniques [Zhang 2011, Tamaru 2014, Ivanov 2014]. In order to realize these gains, however, these setups have been either narrowband [Zhang 2011, Ivanov 2014] or required readjustment at each measurement frequency [Tamaru 2014].

We have developed a Michelson-type microwave interferometer (MMI) for use in VNA-FMR experiments. The MMI is a passive, two-port network consisting of a planar cross junction in a coplanar waveguide transmission line with two short-terminated arms of identical length, as presented schematically in Fig. (a). The nominal characteristic impedance of all transmission lines in the device is 50 Ω, and the center conductor width is 360 μm. At discrete frequencies determined by the arm length $L$, the reflected wave from the two arms destructively interfere at port 1 or port 2 as schematically shown in Fig. 1(b). At these frequencies measurements of very high sensitivity become possible, drastically reducing the measurement time.

The scattering properties of a symmetric planar cross junction can be easily understood in terms of simple nodal analysis based on Kirchoff's Laws, under the assumption that the feature dimensions of the cross junction are much smaller than the wavelength of the microwaves employed. If $E_i$ is the scalar value of the electric field incident on one arm of the cross junction, $E_b$ is the backscattered electric field that propagates back into the arm of the incident electric field, and $E_f$ is the forward scattered electric field in the three other arms, then node analysis requires that $E_f = -E_b = E_i/2$. Based on this fundamental result, we can then calculate the transmitted and reflected electric field amplitudes in the case where two of the cross-junction legs are terminated with shorted transmission lines of length $L$. In that case, the complex value of the transmitted electric field $\tilde{E}_t$ a distance $x$ from the cross-junction along the remaining un-terminated leg is given by

$$\tilde{E}_t = \frac{1}{2} E_i \left(1 - e^{i2\beta L}\right) e^{i\beta x} \tag{1}$$

where $\beta \doteq 2\pi n_{\text{eff}} f/c$ is the transmission line phase constant, $n_{\text{eff}}$ is the effective refractive index of the transmission line, $f$ is the frequency, and $c$ is the speed of light in vacuum. We therefore see that the transmission from port P1 to port P2 is nulled when the length of the terminated junctions is chosen such that $\beta L = n\pi$ where $n$ is an integer. Similarly, the backscattered field is given by

$$\tilde{E}_b = -\frac{1}{2} E_i \left(1 + e^{i2\beta L}\right) e^{i\beta x} \tag{2}$$

Hence, the reflection is nulled for the complementary condition of $2\beta L = (2n+1)\pi$.

To measure FMR with such an interferometer, we can place a sample onto one of the interferometer arms with a shorted termination such that the sample is excited by the ac magnetic fields proximate to the waveguide as sketched in Fig. 1(a). We then make use of the fact that the inductive voltage generated in the waveguide by the magnetization precession in the sample under the condition of ferromagnetic resonance is antisymmetric with regard to direction of propagation [Silva 1999]. This antisymmetric property results in an opposite sign for the backscattered and forward scattered waves radiating from the sample. The forward scattered wave is shifted in phase by 180 degrees upon reflection from the shorted termination at the end of the transmission line. Thus both inductive signals constructively interfere when they arrive back at the cross junction.

Experiment

We have fabricated such a device, and characterized its performance by use of a two-port VNA. The magnitude of the measured complex $S_{11}$ and $S_{21}$-parameters of the MMI are plotted in Fig. 1. We observe a set of notches with a frequency spacing of approximately 400MHz. At the notch frequencies the signals from the two arms interfere destructively, strongly suppressing the amount of power at Port 2 of the VNA. From the cancellation conditions for the reflected and transmitted electric field amplitudes, this spacing agrees with the chosen arm length of L=0.25 m. The degree of the background suppression is a function of both the phase and amplitude imbalance between the two interferometer arms, which we have plotted in Fig. 3.

To measure FMR, we place the ferromagnetic (FM) film on one of the two arms and apply an external magnetic field $H_0$ as shown in Fig. 1(a). The film is Ta(3)/Ni$_{80}$Fe$_{20}$(5)/Ta(3) grown by sputter deposition on an oxidized silicon substrate, where numbers in parentheses are the nominal layer thickness in nanometers. In order to optimize the phase and amplitude balance of the interferometer, a bare coupon of the substrate that has identical dimensions as the FM-covered substrate is placed on the reference arm, thus compensating the dielectric response of the common substrate material. The power applied at port 1 is 0 dBm in all measurements. No trace averaging was utilized, and the intermediate frequency (IF) bandwidth is set to 500 Hz.

We fix the frequency to the notch in $S_{21}$ (i.e. the transmitted signal) at 1.2682 GHz, and then record the S-parameters as a function of swept magnetic field $H_0$. We plot the $S_{11}$ parameter vs. $H_0$ (i.e. the reflected signal) in Fig. 4(a). This signal does not benefit from the advantage of any background suppression ($S_{11}$(1.2682GHz) = - 2.5 dB). For comparison, we plot $S_{21}$ vs. $H_0$ (i.e. the transmitted signal) in Fig. 4(b). In this case, the background signal is strongly suppressed ($S_{21}$(1.2682GHz) = - 36 dB from Fig. 2). The reduction in noise is evident from inspection of Fig. 4(a) and 4(b).

To determine the signal-to-noise ratio (SNR), we fit the measured curves to the complex $\chi_{yy}$ component of the magnetic susceptibility $\chi$ for an in-plane bias field geometry [Weiler 2014]. The dimensionless signal amplitude $A_{ij}$ for S-parameter $S_{ij}$ is taken as the amplitude of the susceptibility fit to the measured resonance,

$$A_{ij} \doteq \frac{S_{ij}(H_0) - O}{\chi_{yy}(H_0)} \qquad (3)$$

where $O$ is the fitted field-independent offset of the scattering parameter. The dimensionless noise amplitude $N_{ij}$ is defined as the root-mean-square of the residuals between the measured spectra and the susceptibility fit,

$$N_{ij} \doteq \sqrt{\langle\left(A_{ij}\chi_{yy}(H_0) - (S_{ij}(H_0) - O)\right)^2\rangle} \qquad (4)$$

where we have verified that the noise amplitude is independent of magnetic field.

In Fig. 4(c), we plot the SNR defined as $\frac{A_{ij}}{N_{ij}}$ or $20\log\left(\frac{A_{ij}}{N_{ij}}\right)$ in dB of the background-suppressed $S_{21}$ and background-full $S_{11}$ FMR measurements as a function of the number of field-sweep averages in $S_{11}$. As is evident from Fig. 4(d), the background-suppressed measurements exhibit a 20 dB increase in the SNR for a single measurement and the background-full $S_{11}$ measurements require 129 averages to obtain the same SNR as that obtained via the $S_{21}$ measurements. In other words, we observe a factor of 129 speedup of data acquisition rate due to the background suppression of the interferometer.

In order to validate the accuracy of VNA-FMR measurements with the MMI, we measured FMR spectra with a conventional thru-line coplanar waveguide transmission line. By fitting these spectra to the in-plane susceptibility $\chi_{yy}$ we can make a direct comparison with the extracted parameters from the measurements with the MMI and non-interferometric measurements. We plot the result in Fig. 5. Within the limits of the prototype we find good agreement between the two techniques.

Discussion

The SNR improvement in our measurements is a direct consequence of the background-suppression achieved with the MMI. In typical VNA-FMR measurement of thin magnetic films with low-loss coplanar waveguides, the background is much larger than the signal. In such cases, trace averaging can be required to eliminate artifacts due to phase and amplitude drift. While a quantitative calculation of the filling factor for the VNA-FMR geometry – especially for conducting films – is beyond the scope of this work, we are able to calculate the relative change in the S-parameters under certainly simplifying assumptions.

For a transmission measurement in which a uniform excitation field from the coplanar waveguide excites uniform precession, the change in the scattering parameters of the loaded transmission line at the ferromagnetic resonance field $H_{res}$ is given by [Silva 2016]

$$\Delta S_{21}(H_{res}) = \frac{\gamma\mu_0 M_s l d_m}{8 Z_0 \alpha w} \qquad (5)$$

where $M_s$ the saturation magnetization, $l$ is the length of the film along the waveguide, $d_m$ the film thickness, $Z_0$ the characteristic impedance of the transmission line, and $w$ is the width of the center line of the coplanar waveguide. For typical film parameters [Silva 2016], $\Delta S_{21}(H_{res})$ is small, on the order of -50 dB. In other words, the signal power can be as small as 0.001 % of the background power at the receiver. By suppression of the background by a certain amount by use of the MMI, the dynamic range is increased by the same amount until the background is suppressed below the signal level. However, the quantifiable reduction in noise due to suppression of the background will be specific to the VNA.

In conclusion, we have developed a broadband microwave interferometer suitable for ferromagnetic resonance measurements. The interferometer does not require adjustment as the measurement frequency is changed, and we have demonstrated a background suppression of up to 35 dB, and a commensurate SNR improvement of 20 dB for a single field-sweep measurement. Improved sensitivity may be exploited to attain faster measurement times or to measure samples of lower magnetic volume. We have demonstrated a speed-up of a factor of 129 compared to non-interferometric measurements. Significantly, the interferometer is compatible with existing VNA-FMR setups as a drop-in replacement for the coplanar waveguide transmission line, with only minor changes to the measurement protocol. Future work will focus on increasing the bandwidth of the interferometer by lithographic patterning and proper channelization of the coplanar waveguide transmission line [Simons 2004].

Acknowledgements

ERJE acknowledges support from the National Research Council Postdoctoral Research Associates program.

References

[Boone 2013] Boone, C T, Nembach H T, Shaw J M, Silva T J (2013) "Spin transport parameters in metallic multilayers determined by ferromagnetic resonance measurements of spin-pumping," *J. Appl. Phys.*, vol. 113, 153906, doi: 10.1063/1.4801799
[Ivanov 2014] Ivanov E N, Kostylev M (2014) "Extremely high-resolution measurements of microwave magnetisation dynamics in magnetic thin films and nanostructures," arXiv:1402.3459
[Kalarickal 2006] Kalarickal S S, Krivosik P, Wu M, Patton C E, Schneider M L, Kabos P, Silva T J, Nibarger J P (2006) "Ferromagnetic resonance linewidth in metallic thin films: Comparison of measurement methods," *J. Appl. Phys.*, vol. 99, 093909, doi: 10.1063/1.2197087
[Maksymov 2014] Maksymov I S, Kostylev M (2014) "Broadband stripline ferromagnetic resonance spectroscopy of ferromagnetic films, multilayers and nanostructures," *Physica E,* vol. 69, pp. 253-293, doi: 10.1016/j.physe.2014.12.027
[Nembach 2011] Nembach H T, Silva T J, Shaw J M, Schneider M L, Carey M J, Maat S, Childress J R (2011) "Perpendicular ferromagnetic resonance measurements of damping and Landé $g$–factor in sputtered $(Co_2Mn)_{1-x}Ge_x$ thin films," *Phys. Rev.* B, vol. 84, 054424, doi: 10.1103/PhysRevB.84.054424
[Neudecker 2006] Neudecker I, Woltersdorf G, Heinrich B, Okuno T, Gubbiotti G, Back C H, (2006) "Comparison of frequency, field, and time domain ferromagnetic resonance methods," *J. Magn. Mag. Mat.*, vol. 307, pp. 148-156, doi: 10.1016/j.jmmm.2006.03.060
[Pozar 2011] Pozar D M (2011), *Microwave Engineering.* New York, NY, USA: John Wiley & Sons

[Poole 1983] Poole C P (1983), *Electron Spin Resonance: A Comprehensive Treatise on Experimental Techniques*. New York, NY, USA: John Wiley & Sons

[Shaw 2011] Shaw J M, Nembach H T, Silva T J (2011) "Damping phenomena in $Co_{90}Fe_{10}$/Ni multilayers and alloys," *Appl. Phys. Lett.*, vol. 99, 012503, doi: 10.1063/1.3607278

[Shaw 2013] Shaw J M, Nembach H T, Silva T J (2013) "Measurement of orbital asymmetry and strain in $Co_{90}Fe_{10}$/Ni multilayers and alloys: Origins of perpendicular anisotropy," *Phys. Rev. B*, vol. 87, 054416, doi: 10.1103/PhysRevB.87.054416

[Silva 1999] Silva, T J, Lee C S, Crawford T M, Rogers C T (1999) "Inductive measurement of ultrafast magnetization dynamics in thin-film Permalloy," *J. Appl. Phys.*, vol. 85, 7849, doi: 10.1063/1.370596

[Silva 2016] Silva T J, Nembach H T, Shaw J M, Doyle B, Oguz K, O'brien K, Doczy M (2016), "Characterization of Magnetic Nanostructures for Spin-Torque Memory Applications with Macro- and Micro-Scale Ferromagnetic Resonance" in *Characterization and Metrology for Nanoelectronics.* Pan Stanford Publishing Pte. Ltd.

[Simons 2004] Simons R (2001) *Coplanar Waveguide Circuits, Components, and Systems.* New York, NY, USA: Wiley-IEEE Press.

[Tamaru 2014] Tamaru S, Yakushiji K, Fukushima A, Yuasa S, Kubota H (2014) "Ultrahigh Sensitivity Ferromagnetic Resonance Measurement Based on Microwave Interferometer," *IEEE Magn. Lett.*, vol. 5, 3700304, doi: 10.1109/LMAG.2014.2365435

[Weiler 2014] Weiler M, Shaw J M, Nembach H T, Silva T J (2014) "Phase-Sensitive Detection of Spin Pumping via the ac Inverse Spin Hall Effect," *Phys. Rev. Lett.*, vol. 113, 157204, doi: 10.1103/PhysRevLett.113.157204

[Zhang 2011] Zhang H, Divan R, Wang P (2011) "Ferromagnetic resonance of a single magnetic nanowire measured with an on-chip mi- crowave interferometer," *Rev. Sci. Instrum.*, vol. 82, 054704, doi: 10.1063/1.3593502

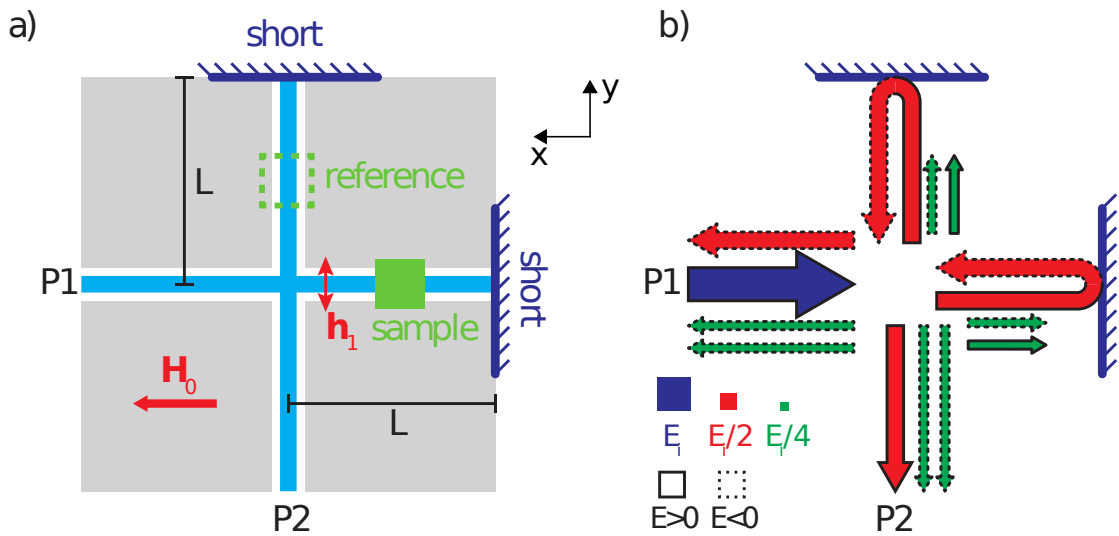

Fig. 1 (a) Schematic of the MMI showing the dc applied field $H_0$, the microwave applied field $h_1$, the positions of the sample and reference, and the interferometer arm length $L$. (b) Schematic depiction of scattering processes in the interferometer for $\beta L = n\pi$. Arrows represent traveling electromagnetic waves. Arrow width/color encodes electric field amplitude and arrow outline encodes sign. The first order scattering process ($E_i \rightarrow \frac{E_i}{2}$, wide to medium wide arrows) results in scattered waves of opposite sign at P1 (backscattering, dashed outline) and P2 and terminated arms (forward scattering, solid outline). At the terminations, these waves are reflected and invert sign. At the cross, the second order scattering process ($\frac{E_i}{2}, \rightarrow \frac{E_i}{4}$, medium wide to narrow arrows) results in two forward scattered waves with amplitude $-\frac{E_i}{4}$ per port. The three outgoing waves constructively (destructively) interfere at Port 1 (Port 2).

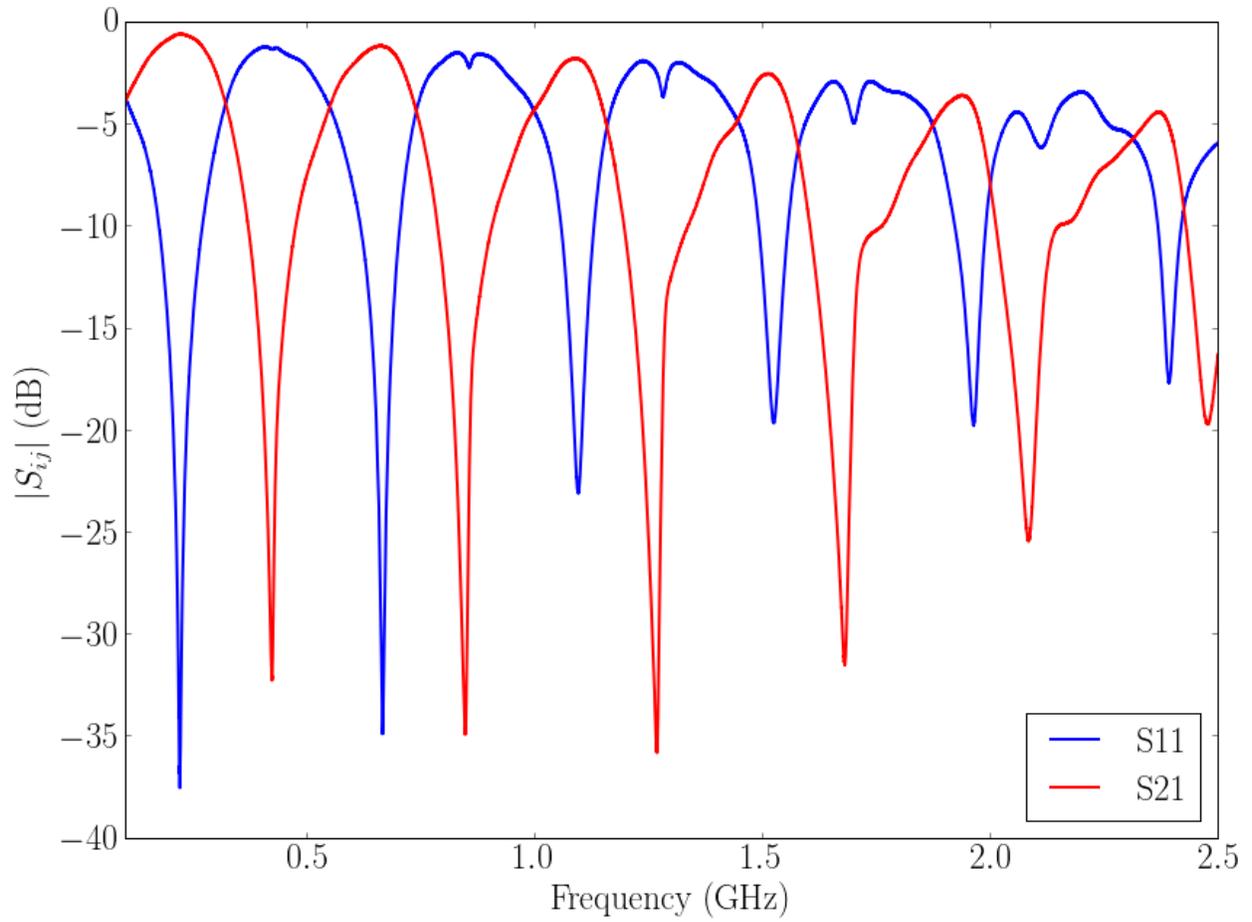

Fig. 1. Magnitude of the complex $S_{11}$ and $S_{21}$ S-parameters of the unloaded MMI in zero external field measured with a two-port VNA.

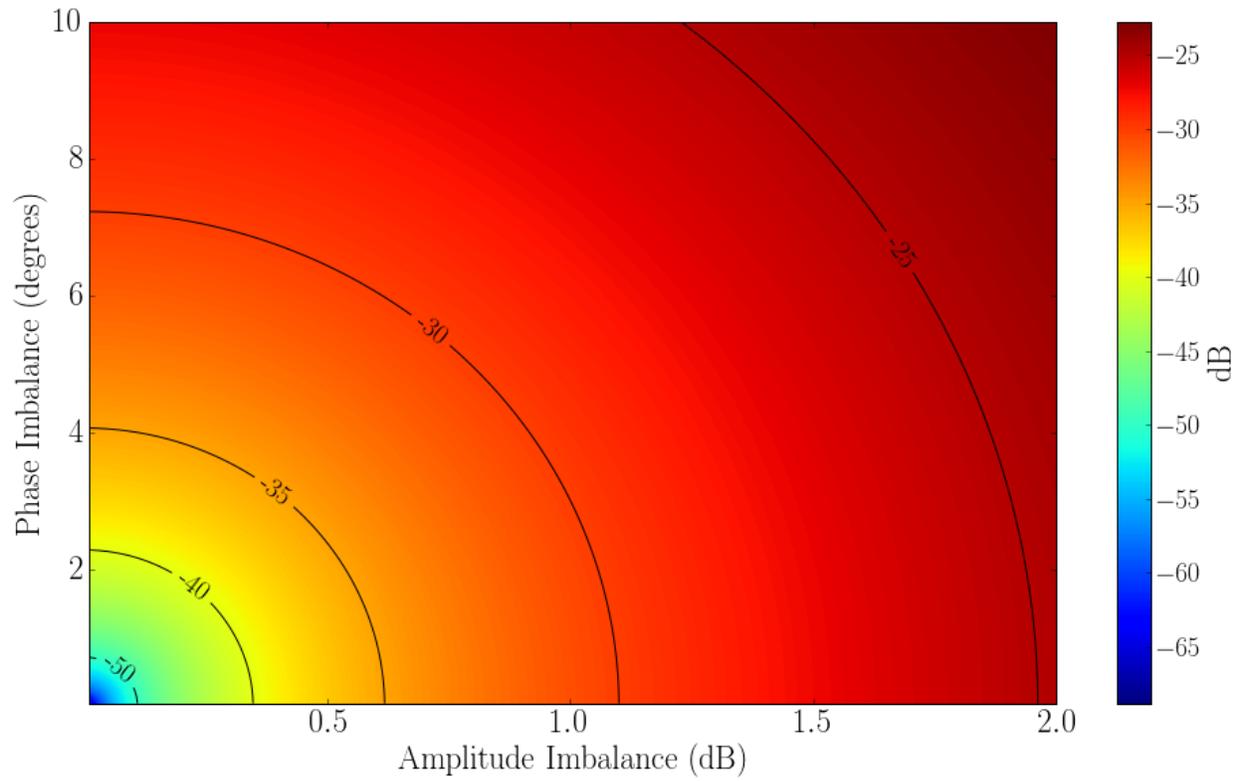

Fig. 3. The amount of background suppression realized from the linear superposition of two waves as a function of the phase imbalance from 180 degrees and amplitude imbalance from equal amplitude given in dB. The black contour lines are labeled with constant values of the suppression in dB.

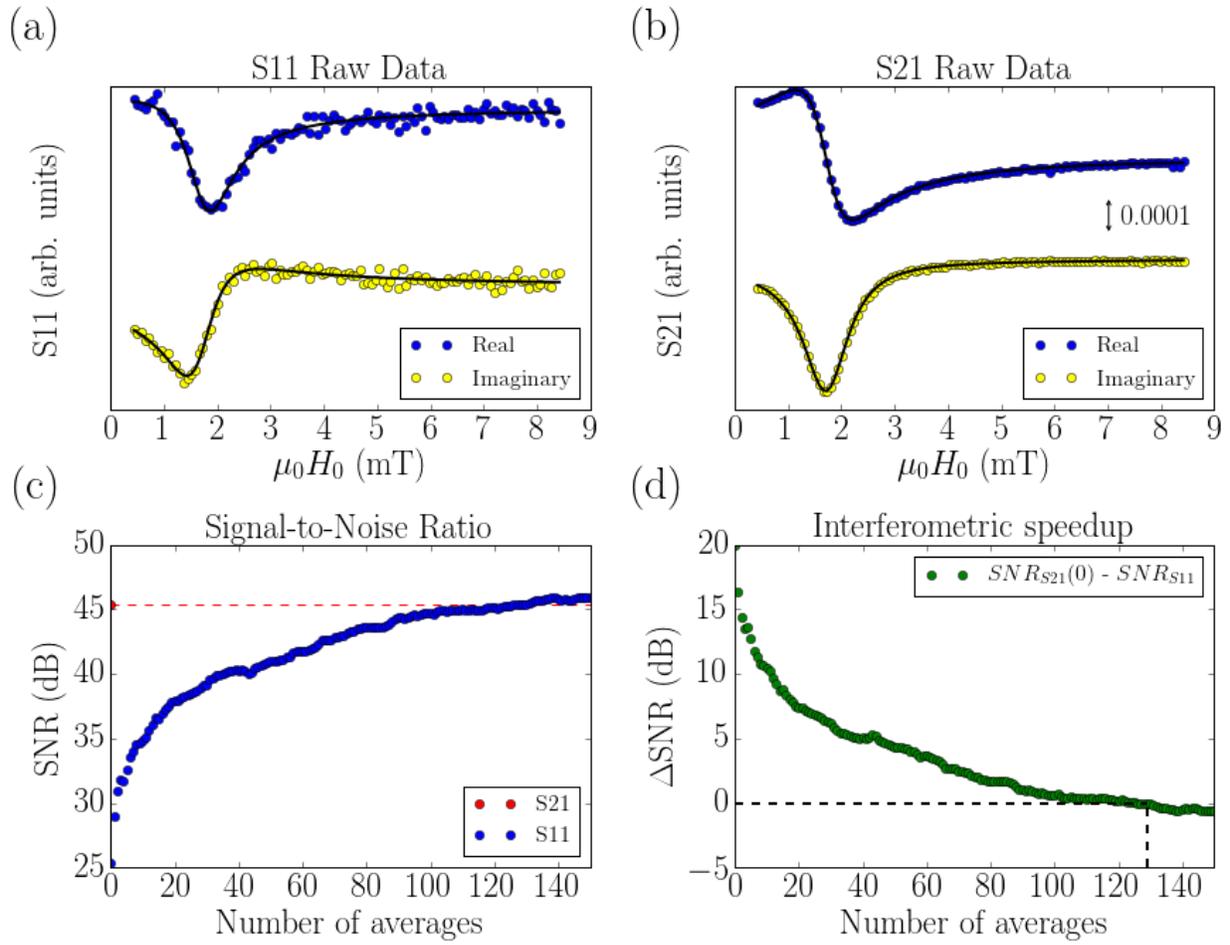

Fig. 4. (a) Sample-loaded $S_{11}$ as measured with a two-port VNA as a function of bias magnetic field. (b) Sample-loaded $S_{21}$ as measured with a two-port VNA as a function of bias magnetic field. (c) The SNR plotted in dB for the number of data averages. The $S_{21}$ SNR is plotted without averaging with the dotted line for comparison to $S_{11}$ data. (d) The difference in the SNR of $S_{21}$ and $S_{11}$ measurements shows that it takes 129 averages of $S_{11}$ data to achieve the unaveraged $S_{21}$ SNR.

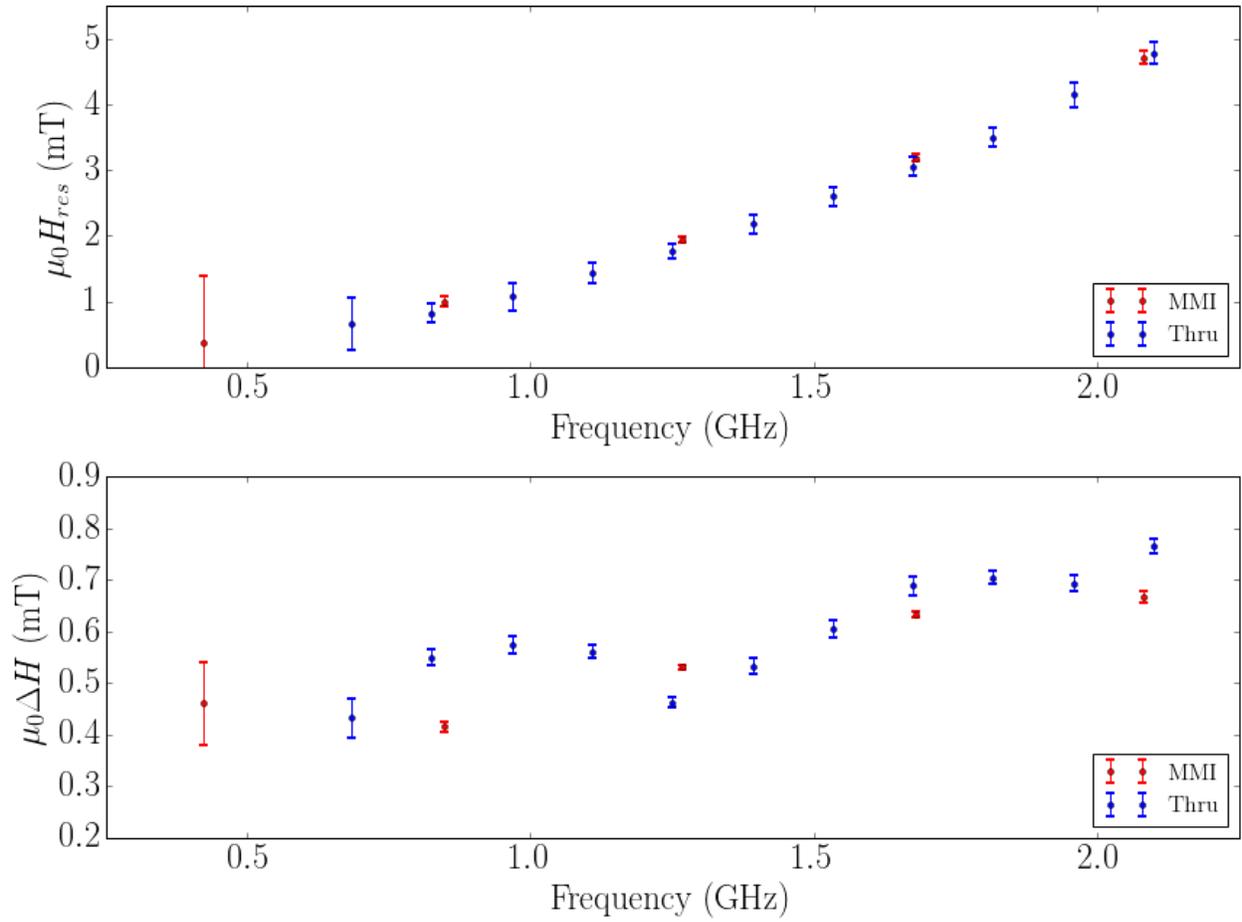

Fig. 5. A comparison of interferometric ("MMI") and conventional ("Thru") FMR parameters extracted from the corresponding spectra.